# ARTICLE

# Wide-Surface Furnace for *In Situ* X-Ray Diffraction of Combinatorial Samples using a High-Throughput Approach

Giulio Cordaro,*[a,b] Juande Sirvent,[c] Cristian Mocuta,[d] Fjorelo Buzi,[c] Thierry Martin,[a] Federico Baiutti,[c] Alex Morata,[c] Albert Tarancòn,[c] Dominique Thiaudière[d] and Guilhem Dezanneau*[a]

The combinatorial approach applied to functional oxides has enabled the production of material libraries that formally contain infinite compositions. A complete ternary diagram can be obtained by pulsed laser deposition (PLD) on 100 mm silicon wafers. However, interest in such materials libraries is only meaningful if high-throughput characterization enables the information extraction from the as-deposited library in a reasonable time. While much commercial equipment allows for XY-resolved characterization at room temperature, very few sample holders have been made available to investigate structural, chemical, and functional properties at high temperatures in controlled atmospheres. In the present work, we present a furnace that enables the study of 100 mm wafers as a function of temperature. This furnace has a dome to control the atmosphere, typically varying from nitrogen gas to pure oxygen atmosphere with external control. We present the design of such a furnace and an example of X-ray diffraction (XRD) and fluorescence (XRF) measurements performed at the DiffAbs beamline of the SOLEIL synchrotron. We apply this high-throughput approach to a combinatorial library up to 735 °C in nitrogen and calculate the thermal expansion coefficients (TEC) of the ternary system using custom-made MATLAB codes. The TEC analysis revealed the potential limitations of Vegard's law in predicting lattice variations for high-entropy materials.

## Introduction

The quest for more efficient materials follows a pattern that changed little over the years, often involving poorly automated steps, such as synthesis, characterizations, and measurements of functional properties, sometimes coupled with ab initio modeling. While this approach has proven effective in generating knowledge and discovering promising formulations, it remains inefficient due to its extremely time-consuming nature and limited coverage of compositions. High-throughput approaches have demonstrated their efficiency in fields such as genomics, health, and pharmaceutics over the past decades and have been recently gained recognition in materials science. The Materials Genome Initiative (MGI), launched in the USA in 2011, aimed to reduce the time-to-market for materials development by a factor of 2. Since then, both experimental and computational aspects of MGI have attracted global attention from academia, government, and industry.

High-throughput calculations have emerged as a powerful tool to rapidly identifying materials with desired properties among vast compositional space, in fields such as photocatalysis,[1] Li-ion batteries,[2] thermoelectrics, or p-type transparent conducting oxides.[3] This approach enabled the creation of huge databases, such as the Materials Project,[4] AFLOW initiative,[5] NOMAD (Novel Materials Discovery),[6] Max (Materials Design at the Exascale)[7], or OPTIMADE (Open Databases Integration for Materials Design)[8].

In parallel to this modeling-centered approach, experimental facilities have been developed to explore the high-throughput synthesis and characterization of materials for specific applications, such as at NIST and NREL.[9] The substantial amount of experimental data generated is made available to the scientific community following the FAIR principles: Findable, Accessible, Interoperable, and Reusable. This approach complements experimental databases COD,[10] ROD,[11] or MPOD,[12] where crystallographic data, Raman spectra, and materials properties are collaboratively produced by the scientific community.

High-throughput experimentation (HTE) is divided into two main categories: (1) automation, which involves parallel or sequential synthesis and characterization of samples, and (2) combinatorial research, which consists of producing single samples containing multiple compositions in patterns or compositional gradients. Combinatorial samples, also known as material libraries, contain information about broad compositional spaces, such as ternary diagrams. Although the concept dates back to 1965,[13] only recent advances in modern characterization tools and computational methods for large datasets have made HTE practical and valueable.[14] Nowadays,

[a.] Université Paris-Saclay, CentraleSupélec, CNRS, Gif-sur-Yvette 91190, France.
[b.] Chimie ParisTech, Université PSL, CNRS, Institut de Recherche de Chimie Paris (IRCP), Paris 75005, France.
[c.] Nanoionics and Fuel Cells group, Catalonia Institute for Energy Research (IREC), Jardins de les Dones de Negre 1, Barcelona 08930, Spain.
[d.] Synchrotron Soleil, Gif-Sur-Yvette 91192, France.

several commercial facilities exist for characterizing combinatorial samples at room temperature, such as XY-resolved X-ray diffraction (XRD), Raman Spectroscopy, UV-Visible spectroscopy. A noteworthy example is the combined X-ray diffraction and fluorescence (XRD/XRF) experiment conducted at the Stanford Synchrotron Radiation Lightsource (SSRL) by J. M. Gregoire *et al.* in 2014 on combinatorial libraries deposited on 100 mm silicon substrates.[15] This setup was optimized for texture analysis at room temperature under ambient conditions. Synchrotron X-ray absorption near edge spectroscopy (XANES) has also been used to study large combinatorial libraries for solar light absorbers in ambient air.[16] However, the *in situ* characterization of combinatorial libraries at high temperatures remains challenging, regardless of the target properties. J. Wolfman's group at Greman laboratory in Tours developed combinatorial procedures using 10 mm side square samples to explore electrical properties of oxides for microelectronics above room temperature (i.e., 400 K).[17] Yet, typical combinatorial libraries are deposited on 100 mm substrates, which complicates measurements due to the lack of commercial equipment for such large sample sizes. A few specialized devices have been developed specifically to address this challenge: (1) MicroXact, a commercial company, offers a probe system for measuring target properties at high temperatures under controlled atmospheres. Their device supports motorized or semi-automated testing of 100 mm or larger wafers up to 700 °C in vacuum or 650 °C at atmospheric pressure.[18] This equipment was adapted at IREC for high-temperature electrochemical impedance spectroscopy (EIS) of 75 mm combinatorial libraries.[19,20] (2) Yan *et al.* developed a thermoelectric screening tool capable of measuring the Seebeck coefficient and electrical resistivity from 300 K to 800 K on 76.2 mm diameter combinatorial thin films.[21] (3) Papac *et al.* demonstrated an instrument for spatially resolved high-temperature EIS of 50 mm libraries.[22]

We recently demonstrated the feasibility of HTE using a combinatorial material library produced by pulsed laser deposition (PLD) and characterized by XY-resolved XRD, XRF, ellipsometry, Raman spectroscopy, high-temperature EIS, and [18]O isotopic exchange depth profiling (IEDP) coupled with time-of-flight secondary ion mass spectroscopy (ToF-SIMS). The experimental data were used as datasets for machine learning routines to predict performance within the $La_{0.8}Sr_{0.2}Co_{1-x-y}Fe_xMn_yO_{3-\delta}$ (LSCFM) ternary system.[20] The experimental and simulated databases are openly available online on GitHub at *https://nanoionicshub.github.io/LSMCF_database/*.

In this study, we present the design of an original furnace for *in situ* XRD and XRF analysis under controlled temperature and atmosphere of 100 mm combinatorial materials libraries. We focus on determining the temperature distribution across the sample and present the XRD results obtained on an LSCFM library as a function of temperature in a nitrogen gas flow. Finally, we evaluated the thermal expansion coefficients for the entire ternary diagram using a high-throughput approach.

## Experimental

### Design and test of the equipment

The furnace was designed to accommodate 100 mm wafers and reach temperatures up to 900 °C under a controlled atmosphere. The schematic is presented in Figure 1, while the technical drawing with orthographic projections is presented in Figure S1 of the Supplementary Information. The supporting structure of the furnace is made of stainless steel and includes a water-cooling circuit. This structure contains two inlets and outlets for gas and cooling-water, as well as a feedthrough for power and thermocouple cables. Temperature control is performed through a 100 mm diameter hotplate that also supports the wafer. To regulate the atmosphere while allowing X-rays to reach the sample, a dome made of PEEK polymer was designed due to its high X-ray transparency.[23] An air-blowing system was included to continuously cool the external surface of the dome and limit its temperature during the experiment. However, this design involving a hotplate suffers from the main drawback of inhomogeneous temperature distribution in the radial direction. This inhomogeneity is further amplified by the requirement to tilt the furnace by 10°, as required by the synchrotron beamline geometry. Therefore, we tested the furnace by using platinum as an internal reference to locally probe the temperature of sample surfaces. First, we established a calibration curve between temperature and the Pt lattice parameter using two reference samples. Then, we partially coated an LSCFM combinatorial library with Pt to serve as an internal probe.

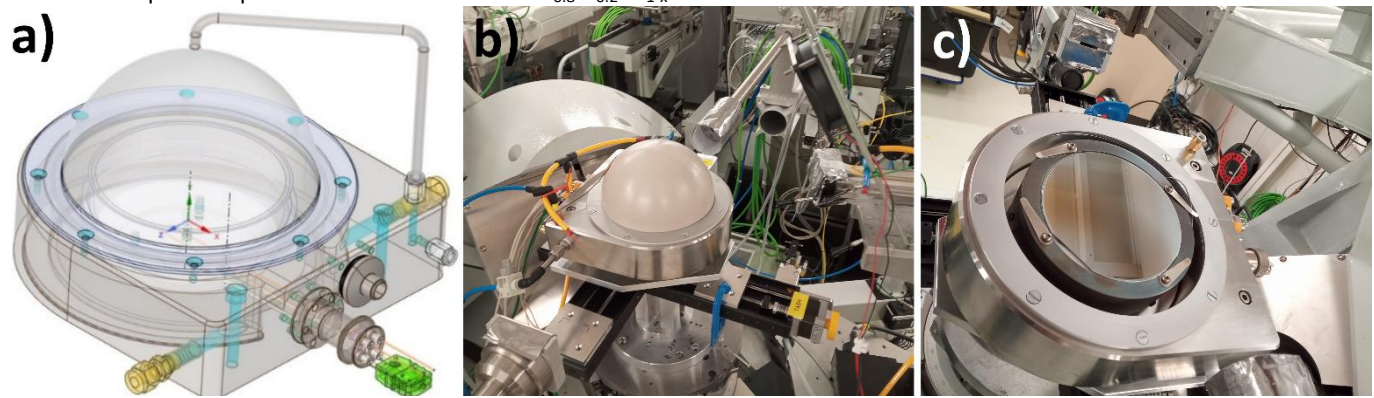

Figure 1 – Scheme (a) and photos (b and c) of the furnace implemented on the DiffAbs beamline at the SOLEIL synchrotron.

Two Pt-coated reference samples were measured to calibrate the actual hotplate temperature. We applied Pt ink to a square single-crystal silicon chip with sides of 10 mm and a circular Si(100) wafer with a diameter of 100 mm. The chip sample was measured in a laboratory-scale diffractometer to obtain an experimental calibration curve of the Pt lattice parameter as a function of temperature. This dependence was subsequently employed to calculate the surface temperature of the Pt-coated wide-surface samples measured inside the furnace. The lab-scale diffractometer was coupled with a Rigaku RA-HF18 rotating Cu anode X-ray generator, operating at 50 kV and 200 mA in a Bragg-Brentano configuration. We measured the Pt-coated wafer in the furnace by collecting 15 maps at 50 °C intervals between room temperature and 735 °C. Then, we acquired 9 maps on the LSCFM combinatorial sample every 100 °C. Finally, the LSCFM library was partially coated with Pt ink as the internal reference, and two high-temperature acquisitions were collected at 400 °C and 700 °C.

**Preparation of combinatorial material libraries**

The material libraries were prepared by combinatorial pulsed laser deposition on Si(100) (*Siegert Wafer*) substrates using a large-area *model 5000 PVD Products* with a 248 nm KrF excimer laser (*Lambda Physik, COMPex PRO 205*). Commercial targets (*Kceracell*) of $La_{0.8}Sr_{0.2}CoO_3$ (LSC), $La_{0.8}Sr_{0.2}FeO_3$ (LSF), and $La_{0.8}Sr_{0.2}MnO_3$ (LSM) were employed to deposit the $La_{0.8}Sr_{0.2}Co_{1-x-y}Fe_xMn_yO_{3-\delta}$ (LSCFM) thin film libraries. The depositions were performed at 700 °C in 0.007 mbar of pure $O_2$ using an ablation frequency of 10 Hz, a laser fluence of approximately 0.8 J cm$^{-2}$, and a target-substrate distance of 90 mm. More details can be found in our previous work.[20]

***In situ* X-ray diffraction/fluorescence (XRD/XRF) experiments and data analysis**

XY-resolved structural and elemental characterizations were performed using synchrotron radiation at the DiffAbs beamline of the SOLEIL facility. X-ray diffraction and fluorescence (XRD/XRF) measurements were conducted using a primary beam energy of 12 keV and an incident angle of 10° between the source beam and the sample surface. The beam size was square with 0.3 mm sides, but, due to the incident angle, the projected area onto the sample surface increased to 1.25 mm in the vertical direction. The output signals were simultaneously collected with an XRD 2D arc detector that covers a range of 135° 2θ (the CirPAD detector[24]) and with an XRF 4-element silicon drift detector placed at a distance of 50 cm. The acquisition was performed within a square mesh coordinate system with sides of 110 mm and lateral steps of 1 mm in both horizontal and vertical directions. The XRD and XRF spectra were collected with an integration time of 0.2 s, and each entire map acquisition took approximately 40 minutes. A total of 12 100 XRD and XRF patterns were obtained, with about 8000 focused on the sample. To enhance the signal/noise ratio of the data, binning with 3 mm intervals was applied, resulting in about 900 patterns to analyze. Experiments were performed in pure nitrogen with a gas flow equal to 3 L h$^{-1}$. The temperature was increased from room conditions (≈ 20 °C) up to a maximum setpoint of 735 °C. The temperature variation is not expected to affect the fluorescence response, which was used to identify the composition at each XY sample position, as well as to verify the stability of combinatorial layers. Therefore, this work focuses on XRD analysis, while XRF results are employed to convert XY coordinates into compositions.

The XRD data were processed to obtain a map of the lattice parameter distribution for each acquisition. First, a threshold calibration of the signal was performed to obtain a uniform response from all the pixels, chips, and modules constituting the detector. A flat-field correction was applied to fix the nonlinear response of each pixel and thus reduce the data noise and dispersion. Finally, geometrical corrections were calibrated using a rotating capillary sample of NIST-certified $LaB_6$ powder as reference.[24] Once the diffractograms were corrected, a MATLAB routine was developed to subtract the background[25], analyze them, and calculate the lattice parameter. In each XRD pattern, all the peaks within the 5-140° 2θ range were identified and fitted with a pseudo-Voigt profile function to collect the angular positions ($2\theta_{Fitted,i}$). The *lsqnonlin* minimization function calculated the cubic lattice parameter ($a_{Cubic}$) and the sample displacement ($SD$) through the following equation:

$$\text{Min function} = \sum_{i=1}^{n°Peaks} \left( 2\theta_{Fitted,i} - 2\sin^{-1}\left(\frac{\lambda \cdot \sqrt{h_i^2 + k_i^2 + l_i^2}}{2 \cdot a_{Cubic}}\right) + \frac{180}{\pi} \cdot \frac{SD}{r_{CirPAD}} \cdot \frac{\sin(2\theta_{Fitted,i})}{\sin(\omega)} \right)^2 \quad (\text{Eq. 1})$$

Where $\lambda$ is the X-ray energy wavelength, $\omega$ is the incident angle, $r_{CirPAD}$ is the goniometer radius (i.e., the sample-detector distance), and $h_i\ k_i\ l_i$ are the Miller indices of each $i$ reflection. The sample displacement ($SD$) is multiplied by $\sin(2\theta_{Fitted,i})$[26] because the synchrotron diffractometer has an asymmetric configuration with a fixed X-ray primary beam and a 2D arc detector. For the lab-scale diffractometer in the Bragg-Brentano configuration, the $SD$ was instead multiplied by $\cos(\theta_{Fitted,i})$.[27] This MATLAB routine was employed to obtain a Pt lattice parameter from each diffractogram collected on the reference chip and wafer, as well as on the partially Pt-coated LSCFM combinatorial library. Using a primary beam energy of 12 keV, we collected up to 16 reflections of the Pt cubic structure (*Fm-3m*, space group n° 225). The same routine was used for LSCFM reflections, changing the space group to the cubic *Pm*-3*m* (n° 221) and the associated Miller indices. The first 12 peaks were fitted and used to calculate the cubic lattice parameter of the perovskites at room temperature. For high-temperature measurements using the dome, only 4 reflections were considered, i.e., (2 0 0), (2 1 1), (2 2 0), and (3 1 0), due to reduced intensity or overlapping with PEEK ones. The massive amount of data required the development of a robust and efficient code to analyze the diffractograms and obtain reliable lattice parameters and temperature values in a reasonable amount of time.

To enhance the accuracy of the temperature values calculated for the Pt-coated wide-surface samples, we experimentally calibrated the Pt thermal expansion curve using the Pt-coated Si chip in the lab-scale diffractometer. The diffractograms were collected in the 30-130° 2θ to include the first 8 reflections of the *Fm*-3*m* Pt structure, from room temperature to 850 °C every 50 °C during heating. While cooling, measurements at 600 °C, 400 °C, 200 °C, and 50 °C were repeated to verify reproducibility. Particular care was taken in temperature collection to minimize the uncertainty. Each XRD measurement was performed at a fixed temperature after 15 minutes of stabilization to ensure complete homogenization of the chamber and, therefore, excellent precision of the registered value and the entire calibration curve. The Pt lattice parameter was calculated through a modified version of the MATLAB code that used a fitting function composed of two pseudo-Voigt profiles, simulating the K$\alpha_1$ (1.5405929 Å) and K$\alpha_2$ (1.5444274 Å) Cu emission lines of the anode tube. A quadratic function was fitted to the experimental points to obtain the direct dependence of the Pt lattice parameter on the temperature. This quadratic function was applied to calculate the surface temperature of large-area samples with Pt ink as the internal reference.

## Results and discussion

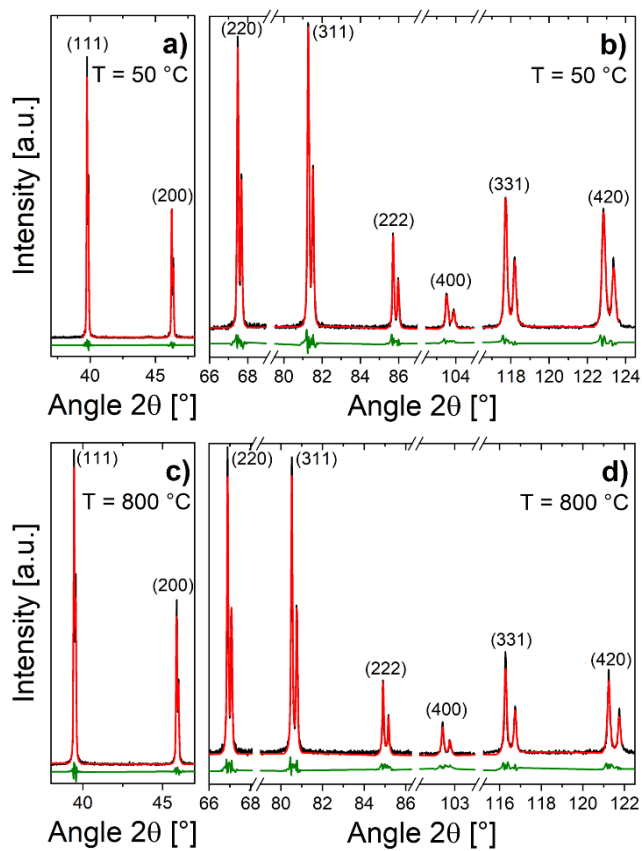

Figure 3 – XRD diffractograms collected on the Pt-coated Si(100) chip in the lab-scale Bragg-Brentano diffractometer with Cu anode tube at 50 °C (a and b) and 800 °C (c and d) with experimental data (black line), fitted curves (red line), and the residuals (green line). The Miller indices of the first 8 reflections are reported. The scale is different among all 4 panels.

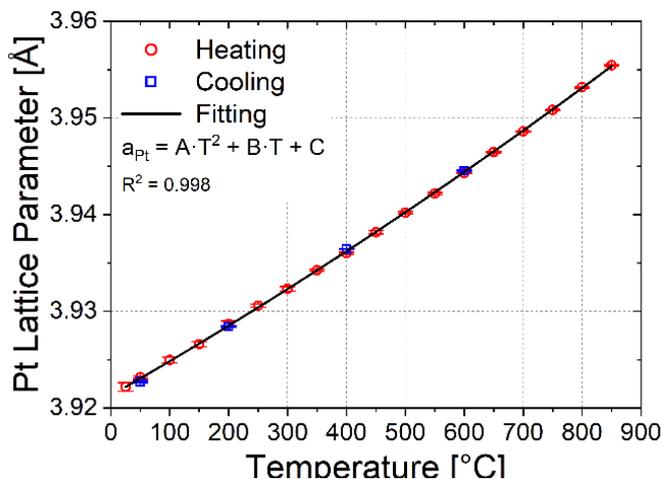

Figure 2 – Calibration of the platinum lattice thermal expansion with calculated unit cell parameters (red squares for heating and blue circles for cooling measurements) and the fitted quadratic curve (line) with the corresponding coefficient of determination ($R^2$).

### Calibration of temperature with the Pt chip sample

As mentioned previously, the hotplate heating system intrinsically leads to an inhomogeneous temperature distribution. This temperature distribution is even more challenging to address when the furnace must be inclined due to the beamline experimental geometry. To solve this issue, *i.e.*, to determine the temperature at each substrate point, we used both Pt external and internal reference samples for probing the temperature. We chose to use platinum because of the excellent chemical and structural stability of its cubic lattice at high temperatures.

First, we collected a calibration curve by measuring the Pt lattice parameter as a function of temperature on a Pt-coated square single-crystal silicon chip. This sample is representative of the wide-surface Pt-coated wafer and the substrate used for the LSCFM combinatorial sample. The Si chip with the Pt layer enabled the calculation of a quadratic function to convert any Pt lattice parameter into temperature within the 50-850 °C range. The diffractograms collected at 50 °C during cooling and 800 °C during heating are reported in Figure 2. The angular range was selected to include the first 8 reflections of the Pt cubic cell using the Cu anode tube. The peak splitting is related to the K$\alpha_1$ and K$\alpha_2$ Cu emission lines, which are considered in the fitting model. The excellent agreement between experimental data (black lines) and the fitting results (red lines) allowed for the extraction of precise values of peak positions, which were used to calculate the Pt lattice parameters with their standard deviations (**Error! Reference source not found.** S1 of the Supplementary Information).

The difference between the heating and cooling values of the Pt lattice parameter was normalized over the heating $a_{Pt}$ to quantify the reproducibility. The percentage difference at 50 °C is equal to 0.012%, while it decreases at higher temperatures, being 0.008%, -0.010%, and -0.008% at 200 °C, 400 °C, and 600 °C, respectively. These tiny discrepancies confirm the reliability of the procedure. Both the Pt cell parameters obtained during heating and cooling, along with their standard deviations, were included in the fitting to calculate the coefficients of the quadratic function of the temperature (Figure 3). The equation with the fitted coefficients A, B, and C is as follows:

$$a_{Pt} = 5.9172 \times 10^{-9} \times T^2 + 3.1914 \times 10^{-5} \times T + 3.9121 \text{ Å (Eq. 2)}$$

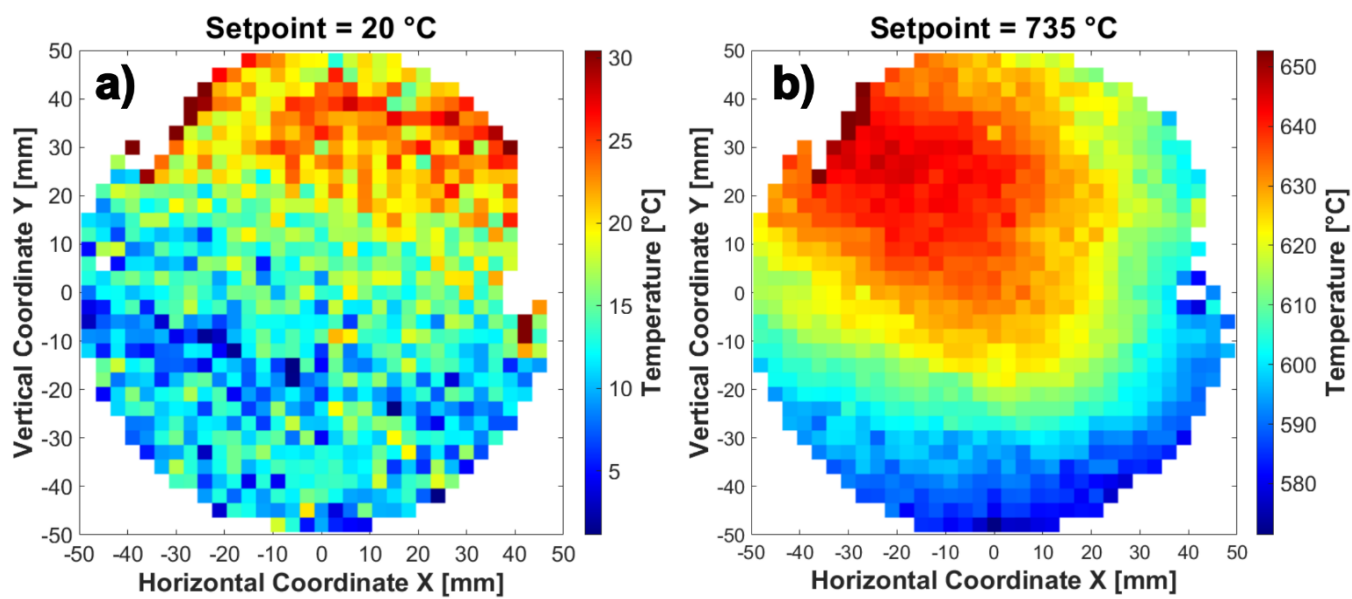

Figure 4 – Heatmap of temperature values calculated from the Pt lattice parameters of the XRD maps collected at room temperature (a) and using a hotplate setpoint of 735 °C (b).

Where T represents the temperature in K degrees.
The formula was inverted to calculate the temperature from the Pt lattice parameter, and the fitting was repeated to improve accuracy. The resulting equation is as follows:

$$T = \left(\sqrt{1.7340 \times 10^8 \times a_{Pt} - 6.7058 \times 10^8} - 2.7875 \times 10^3\right) - 273.15 \text{ °C (Eq. 3)}$$

**Furnace setup in the DiffAbs beamline**

The first sample measured in the furnace was the Pt-coated 100 mm diameter silicon wafer. The furnace was tested at up to 735 °C to limit the dome temperature, which was measured by placing a thermocouple in contact with the external surface. The maximum dome temperature recorded was 170 °C for the 735 °C hotplate setpoint, and we chose to limit the high-temperature run at this value as a precaution. However, higher setpoints could be achieved because the maximum temperature at which PEEK can maintain its mechanical properties without degradation is 250 °C (i.e., continuous service temperature[28]). At each temperature, an XY-resolved collection of XRD diffractograms was performed using an acquisition time of 0.2 s every mm. A continuous *flyscan* was carried out in the horizontal direction to suppress the deadtimes, while 1 mm steps were taken vertically inside a 110 mm square area. Data reconstruction and binning were performed to obtain a diffractogram every 3 mm in both directions. Therefore, we were able to acquire spatially resolved XRD maps of the wide-surface samples at each temperature. Examples of diffractograms with peak fittings are reported in Figure S2 of the Supplementary Information.

The map collected at room temperature allowed for the calibration of the energy shift of the primary electron beam and the assessment of the precision of our procedure. At 12 keV, the energy shift was calculated as -0.046 keV to match the Pt lattice parameter of the center of the sample with the value obtained with the lab-scale measurement (*i.e.*, 3.9222 Å). This energy shift value agrees with the -0.043 keV obtained from the experimental calibration performed using a NIST standard LaB$_6$ reference sample.

At room temperature, the distribution of the calculated temperature values reflects the variations due to uncertainties in the entire calibration process. This process includes the errors of the calibration measurement on the chip presented in the previous section, mainly due to the uncertainty of the temperature recording by the thermocouple placed as close as possible to the sample. Additional inaccuracy is introduced by the intrinsic uncertainties of the synchrotron experimental procedure and the subsequent data treatment. The mean $a_{Pt}$ value obtained at room temperature is 3.9219 Å, and its average standard deviation is 0.0004 Å. This accuracy may appear remarkably small, but when converted into temperature using Equation 3, the resulting range is 19.1 ± 11.7 °C. The standard deviation is very homogeneous across the sample. Therefore, a range of about 10 °C is the best precision we could achieve with our procedure. It is worth noting that the mean standard deviation decreases when the temperature increases, reaching a minimum value of 0.0003 Å, corresponding to 616.6 ± 7.7 °C for the measurement with the hotplate setpoint of 735 °C. This reduction is due to the crystallinity enhancement of Pt deposition and the subsequent increase in peaks' intensity and reduction in peaks' width with temperature increase. Future improvements will be introduced in both data collection and treatment to reduce temperature uncertainty further. Figure 4 presents the temperature heatmaps at room conditions (panel a) and 735 °C (panel b).

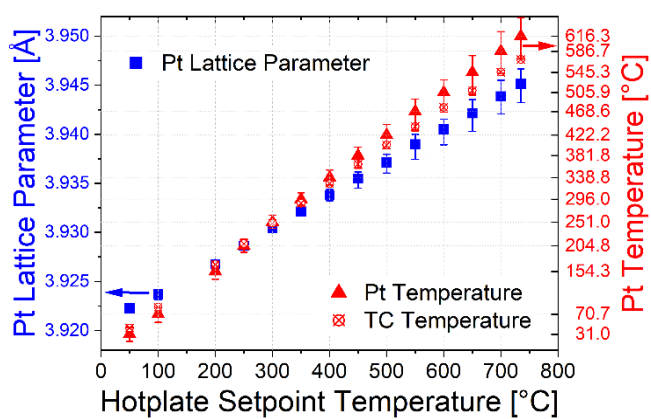

Figure 5 – Pt lattice parameters (blue squares) and corresponding temperatures (red triangles) calculated from XRD maps at different hotplate setpoints. Each symbol represents the mean value, while the bar shows each map's minimum and maximum. Thermocouple temperature values are also reported (red crossed circle).

At high temperatures, the presence of a hot spot is evident. A central thermal maximum is typical of the hotplate heating systems, but we observed that the maximum temperature is in the top left region of the sample. This shift of the hot spot is related to (i) the clamp used to fix the sample, which induces better contact between the sample and the hotplate, and (ii) the 10° tilting of the sample holder, which moves the heat convection towards the top due to gas recirculation inside the dome. The irregular shape of the temperature distribution and its dependence on the clamping limit the applicability of this calibration process for another sample. It could lead to significant errors if any Pt map is used as an external temperature probe without verification each time. Therefore, it is evident that an internal probe will always be necessary for a combinatorial sample. This result motivated us to partially cover the LSCFM combinatorial library with Pt.

At room temperature (Figure 4a), the values are randomly distributed, except for the top part of the sample, which presents an average increase of 0.0002 Å compared to the bottom area. This increase in $a_{Pt}$ is responsible for an average of 4.5 °C larger values and is related to the reconstruction of the diffractograms. Due to the detector structure, the diffractograms present 19 "dead zones" corresponding to the overlap of different modules composing the CirPAD detector.[24] Therefore, when a Pt peak is close to one of these "dead zones", the standard deviation of the fitting increases, and its relative importance in the minimization routine is reduced. Removing the inverse of the standard deviation as the weighting factor improves precision but significantly reduces accuracy. We therefore retained the fitting weights, at the expense of minor oscillations. These variations are negligible at higher temperatures (Figure 4b).

An additional source of uncertainty could be due to the sample displacement calculation (Eq. 1). We tested different formulas with more parameters for the sample displacement, but the precision did not improve. For example, we added a constant term, or a $\cos(\theta_{Fitted,i})$ factor as in the Bragg-Brentano configuration, or a combination of these. The $\sin(2\theta_{Fitted,i})$ dependency gave the best result. We also performed an additional verification by varying the vertical displacement of a LaB$_6$ standard plate sample to evaluate the lattice parameter variation. Eq. 1 calculated the LaB$_6$ lattice parameters equal to 4.154(9) Å with a maximum standard deviation of 0.0002 Å in between -1 and +1 mm of vertical displacement using $\omega$ = 10°. The sample displacement distributions for the Pt-coated wafer show a constant and monotonic decrease between 0.7 mm and -0.6 mm (not reported), entirely within the range verified by the measurement with the LaB$_6$ standard plate. This vertical displacement of the Pt wafer can be explained by the difficulty of placing the sample surface perfectly parallel to the horizontal axis of the goniometer. For all temperatures, the sample displacement distribution presents the same trend, which is not expected to change. A slight variation of the sample displacement could arise from the thermal expansion of the sample and furnace support; therefore, the sample height was aligned before each measurement collection. The difference in sample height alignment between the lowest and the highest temperatures was 0.3 mm. It is worth mentioning that a fine determination of internal reference peaks should allow for the refinement of the sample displacement without requiring sample height and parallelism alignments.

The results of all the maps of Pt cell parameters and relative temperatures are summarized in Figure 5. The $a_{Pt}$ values are reported as blue squares, while the corresponding temperature values are red triangles. The symbols represent the mean values of the distribution of each map, while the minimum and maximum values are shown as bars below and above each symbol. The range of values of each map increases with the temperature, as expected, due to more significant thermal losses. Table S2 contains all the results. The maximum range is thus obtained at 735 °C, with values spanning from 571.4 ± 13.9 °C up to 652.8 ± 3.9 °C (Figure 4b).

The temperature recorded by a thermocouple is presented as crossed empty circles (Figure 5). The thermocouple tip is placed on the right side of the sample, below one of the two clamps that prevent the sample from moving during the acquisition time. The clamp shadowing effect is responsible for the two regions without Pt results. The right clamp with the thermocouple is thus located in an area with temperatures close to the minimum values.

**Application to the LSCFM combinatorial library**

The results of the *in situ* measurements performed on the LSCFM combinatorial library without the Pt internal reference are presented. The distribution of the LSCFM lattice parameters is reported in the maps at room temperature and 735 °C (Figure 6). The diffractograms of the LSC, LSF, and LSM deposition centers are presented in Figure S2 for 20 °C (without the PEEK dome) and in Figure S3 for 735 °C (using the dome).

The variations in the unit cell size are related to the different composition present on each XY coordinate of the combinatorial sample. An exhaustive high-throughput characterization of the LSCFM ternary system using combinatorial samples can be found in our previous publication.[20] These high-throughput techniques include room temperature XRD, performed with a lab-scale diffractometer, and XRF, carried out at the synchrotron inside the custom furnace, by simultaneously collecting the fluorescence and diffraction signals. XRF results provided a map of the Co, Fe, and Mn molar ratios, normalized into LSC, LSF, and LSM compositions. This map is a fundamental calibration tool for relating the XY coordinates of a combinatorial sample with the layer composition at that point, enabling the following characterization results to be plotted in a ternary diagram (see Fig. 2 of ref.[20]). The distributions of the LSCFM lattice parameters in Figure 6a and b are very similar, demonstrating the thermal resistance of the combinatorial layer at high temperatures without phase decomposition.

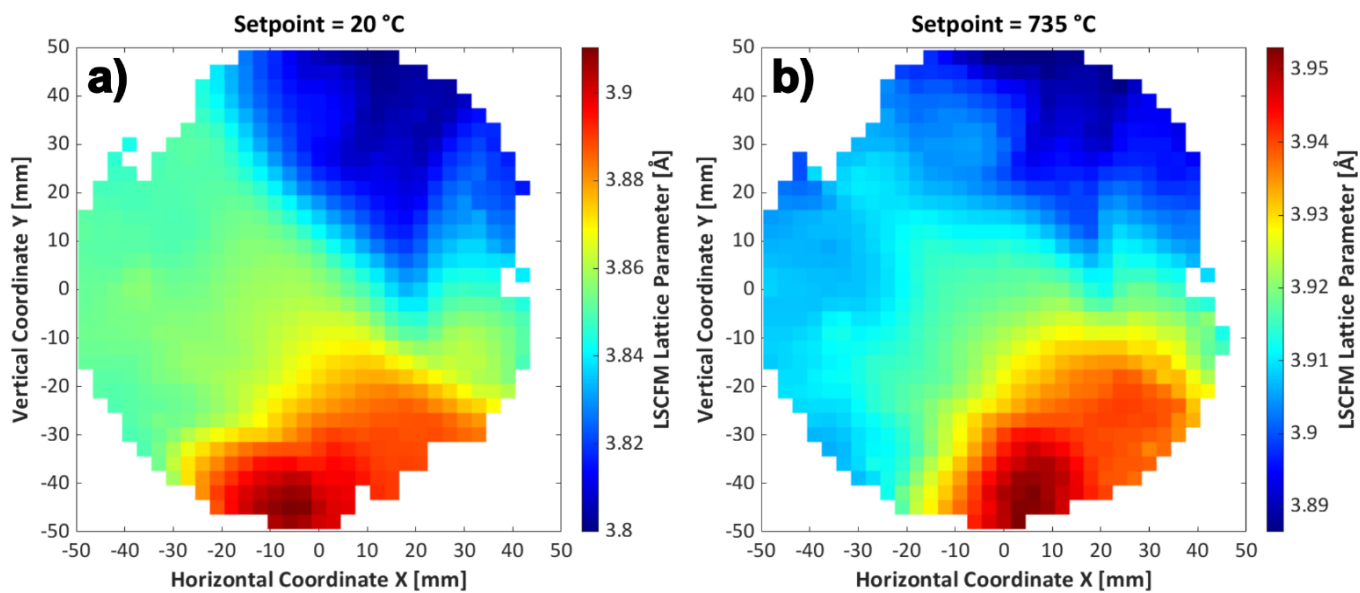

Figure 6 – Heatmap of LSCFM lattice parameters calculated from the XRD maps collected at room temperature (a) and using a hotplate setpoint of 735 °C (b). The $La_{0.8}Sr_{0.2}CoO_{3-\delta}$ (LSC) deposition center is located in the top part of the map and exhibits the lowest lattice parameters; the $La_{0.8}Sr_{0.2}FeO_{3-\delta}$ (LSF) center is in the bottom region, with the highest values; and the $La_{0.8}Sr_{0.2}MnO_{3-\delta}$ (LSM) center is on the left, showing intermediate lattice parameters.

At room temperature, the lattice parameter ranges between 3.7999 ± 0.0014 Å for the Co-rich part and 3.9104 ± 0.0003 Å for the Fe deposition center. These values are slightly different from our previous article but are in line with literature results obtained on LSC powders (pseudocubic $a$ = 3.81-3.85 Å, depending on the oxygen content[29]) and thin films ($a \approx$ 3.826 Å[30]). Also, reference unit cell values for LSF powders (3.900 Å[31]) and thin films (3.896 Å after annealing[32]) are in good accordance with our results. Finally, the LSM lattice parameter presents

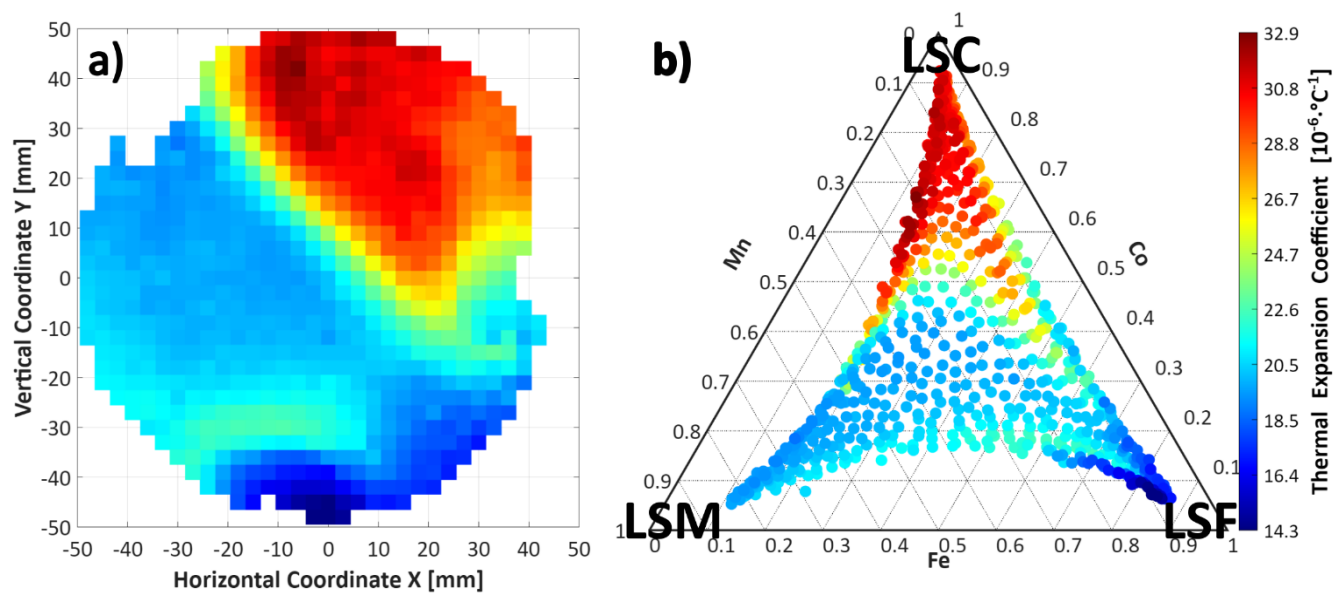

Figure 7 – Heatmap (a) and ternary plot (b) of LSCFM thermal expansion coefficient (TEC) values obtained in nitrogen. The three corners of the ternary diagram represent the pure compositions: $La_{0.8}Sr_{0.2}CoO_{3-\delta}$ (LSC), $La_{0.8}Sr_{0.2}FeO_{3-\delta}$ (LSF), and $La_{0.8}Sr_{0.2}MnO_{3-\delta}$ (LSM). The sides of the triangle correspond to the binary systems (i.e., LSCF, LSFM, and LSCM) and the central area includes all the ternary compositions inside the LSCFM system: the further from a corner, the lesser that compound is present.

intermediate values: 3.843 Å for powders[33] and 3.86 Å for thin films[34].

As expected, the lattice parameters increase at higher temperatures due to thermal expansion. The calculated values (Figure 6b) are in line with the literature: $a_{LSC}$ = 3.90-3.92 Å at 900 °C[35] and $a_{LSF}$ = 3.925-3.962 Å at 600-1000 °C.[31,36]

Collecting 9 entire maps of the LSCFM lattice parameters between 50 °C and 735 °C enabled the calculation of the thermal expansion coefficients (TECs) for each composition inside the ternary system, using the following formula:

$$TEC = \frac{\frac{a_i - a_{50\,°C}}{a_{50\,°C}}}{T_i - 50} \,°C^{-1}$$

Where $a_i$ represents the lattice parameters at temperature $T_i$ expressed in K, and $a_{50\,°C}$ is the lattice parameter at 50 °C. The temperature was corrected using the external calibration of Pt (Table S2) and the internal one of the two additional XRD maps collected at 400 °C and 700 °C after depositing the Pt ink on some areas of the LSCFM combinatorial sample. These Pt internal references allowed for the reduction of temperature uncertainties for TEC calculations. The TEC values range between 13 and 34 × $10^{-6}$ °$C^{-1}$ (Figure 7a). These results are also reported in the ternary plot using the XRF calibration (Figure 7b), where the visualization of TEC distribution is simplified. The Fe-rich area corresponds to the minimum values, which are in good agreement with references ($TEC_{LSF}$ = 13.4 × $10^{-6}$ °$C^{-1}$ [37] or 10.9-14.0 × $10^{-6}$ °$C^{-1}$ [38]). The smallest TECs are expected in this region, as LSF exhibits the largest lattice parameter. Slightly larger values are observed for LSM, while the largest TECs correspond to the Co-rich part, mostly due to the contribution of chemical expansion, i.e., changes in oxygen stoichiometry. This distribution is in line with the literature, but the values obtained for LSC are significantly larger than reference ones: $TEC_{LSC}$ = 19.7-23 × $10^{-6}$ °$C^{-1}$.[39,40] This discrepancy can be explained by the $N_2$ atmosphere used in our measurements, compared to literature experiments performed in air. Previous studies in $N_2$ reported a TEC equal to 34.2 × $10^{-6}$ °$C^{-1}$ for $La_{0.7}Sr_{0.3}CoO_{3-\delta}$ in the 500-600 °C range.[41] Reducing oxygen partial pressure increases the lattice expansion of $La_{1-x}Sr_xCo_{1-y}Fe_yO_{3-\delta}$ (LSCF) because oxygen losses induce a chemical expansion in addition to thermal one.[42–44] This effect is more evident for Co due to its lower melting point and weaker bond strengths.[45,46] Additionally, intrinsic uncertainties are present in our measurements along with the pseudocubic approximation applied to the typical rhombohedral structure of LSC and LSCF powders. The addition of a Pt internal reference and the production of fully polycrystalline layers will strongly reduce the uncertainties related to our high-throughput procedure. However, these results demonstrate the ability to calculate semi-quantitatively the thermal expansion of an entire ternary system using a single combinatorial sample. The internal comparison within this compositional space can be performed using high-throughput experimentation (HTE) as an initial screening to select the most promising compounds.

Furthermore, deeper insight into TEC results involves the verification of Vegard's law inside the ternary system. By fixing the concentration of one cation, it is possible to evaluate the lattice expansion as a function of the ratio between the other 2 elements. Figure 8 shows TEC values varying Co/Fe ratio at fixed Mn contents (panel a), Co/Mn ratio (panel b), and Fe/Mn ratio (panel c). For low Mn amount (Mn = 0.1, i.e., blue circles in Figure 8a), the TEC variation is linear within the 0.1-0.8 Fe range, following Vegard's law. Increasing Mn content (red and orange circles), a linear trend in TEC is observed within a smaller Fe/Co range: Fe = 0.2-0.5 for Mn = 0.2, and Fe = 0.2-0.3 for Mn = 0.3. A similar behavior is found when the Fe amount is constant, and the variable cations are Mn/Co (panel b of Figure 8). Generally, with Co > 0.4, the linearity is easily observed due to high TEC values. However, it is worth noting that for fixed Co amounts, linear behaviors are observed when the Mn/Fe ratio is unbalanced. For example, for Co = 0.2 (blue circles in panel c), two linear regions are possibly present for low (Mn = 0.1-0.3) and high Mn contents (Mn = 0.3-0.6). Increasing the Co amount, a region with constant TEC can be observed in the center of the curves (orange and yellow circles in Figure 8c).

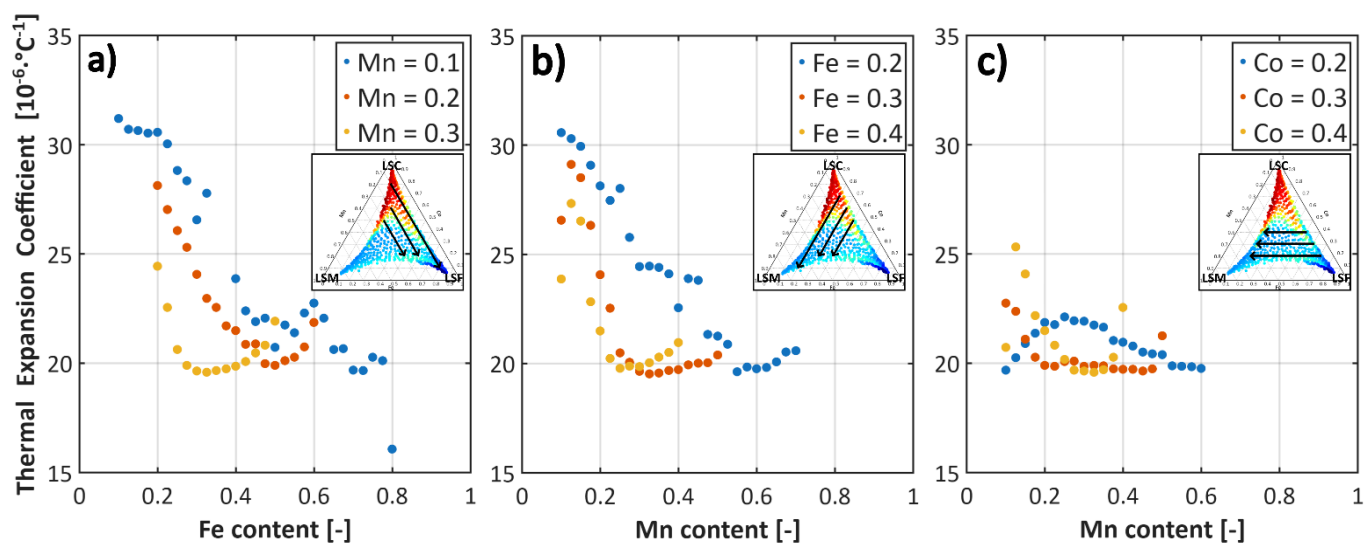

Figure 8 – Thermal expansion coefficients (TEC) values as a function of cation ratio by fixing one element at a time: a) fixed Mn content, Fe-Co ratio varies, b) fixed Fe content, Mn-Co ratio varies, c) fixed Co content, Mn-Fe ratio varies. The cation ratios reported on the x axes are showed using black arrows inside the ternary diagrams in the insets.

Our results suggest that introducing 3 cations in the B site with similar concentrations reduces variations in lattice expansion. A possible explanation for this phenomenon is the enhanced lattice stabilization generated by the increase in entropy for the compositions in the central area of the ternary diagram. The closer the amounts of Co, Fe, and Mn are present in LSCFM, the higher the entropy of the compound is. The highest entropy material is indeed $La_{0.8}Sr_{0.2}Co_{1-x-y}Fe_xMn_yO_{3-\delta}$, with x and y = 0.33, which is the central point of the ternary plot. Due to the "*cocktail effect*", this composition is expected to be the most resilient to compositional variations because it is the furthest from the single compounds (LSC, LSF, and LSM). Therefore, it is possible that high-entropy materials could not follow Vegard's law, but further verification is required.

## Conclusions

We developed a high-throughput experimentation (HTE) procedure for measuring the thermal lattice expansion of an entire ternary diagram for a thin-film combinatorial library. We designed and assembled a custom furnace for *in situ* XRD/XRF measurements performed at the DiffAbs beamline of the SOLEIL synchrotron. By optimizing the data collection and processing parameters, it is possible to identify structural features (e.g., thermal expansion coefficients, phase transitions, water vapor incorporation, and other phenomena) inside a vast thermal, atmospheric, and compositional space. An internal reference is required for precise temperature identification (with a temperature standard deviation < 10 °C), due to the intrinsic radially inhomogeneous temperature distribution typical of hotplates. Cubic lattice parameters can be identified with standard deviations in the order of 0.001 Å, but improvements can be achieved using polycrystalline thin films. Additionally, the diffractogram reconstruction procedure can be further refined.

The developments presented in this work are openly available to the HTE community, specifically the furnace schematics, the codes for processing large datasets, and the combinatorial LSCFM results. The analysis of these results raised the following question: Do high-entropy materials always follow Vegard's law?

## Author contributions

Giulio Cordaro: Conceptualization, Data curation, Formal analysis, Investigation, Software, Visualization, Writing – original draft. Juande Sirvent: Investigation. Cristian Mocuta: Data curation, Formal analysis, Software. Fjorelo Buzi: Investigation. Thierry Martin: Resources. Federico Baiutti: Conceptualization, Supervision, Validation, Writing – review & editing. Alex Morata: Supervision. Albert Tarancòn: Conceptualization, Funding acquisition, Supervision. Dominique Tiaudière: Conceptualization, Investigation, Methodology. Guilhem Dezanneau: Conceptualization, Funding acquisition, Investigation, Methodology, Supervision, Validation, Writing – review & editing.

## Conflicts of interest

There are no conflicts to declare.

## Data availability

Data for this article, including XRD spectra collected on Pt and combinatorial LSCFM thin films between room temperature and 735 °C are available at the *Recherche Data Gouv* at https://doi.org/10.57745/RVHUP6. The data supporting this article have been included as part of the Supplementary Information.

The code for calculating Pt and LSCFM cubic lattice parameters from single maps of XRD diffractograms can be found at https://github.com/CordaroG/XRD_Sy_Pt-LSCFM. The version of the code employed for this study is permanently available at https://archive.softwareheritage.org/browse/directory/40e3a170a3a035b66a4adc6907cbbdea047bc6d8/.

## Acknowledgements

We dedicate this work to the memory of Dominique de Barros, whose fundamental contributions made the furnace development possible. This work was supported by the French National Research Agency (ANR) through the AAPG2021 - CES50 call (AUTOMAT-PROCELLS project – ANR-20-CE05-0001) and through France 2030 DIADEM PEPR program (HIWAY-2-MAT project – 22-PEXD-0008).

# Wide-Surface Furnace for *In Situ* X-Ray Diffraction of Combinatorial Samples using a High-Throughput Approach

Giulio Cordaro,*[a,b] Juande Sirvent,[c] Cristian Mocuta,[d] Fjorelo Buzi,[c] Thierry Martin,[a] Federico Baiutti,[c] Alex Morata,[c] Albert Tarancòn,[c] Dominique Tiaudière[d] and Guilhem Dezanneau*[a]

[a]Université Paris-Saclay, CentraleSupélec, CNRS, Gif-sur-Yvette 91190, France
[b]Chimie ParisTech, Université PSL, CNRS, Institut de Recherche de Chimie Paris (IRCP), Paris 75005, France
[c]Nanoionics and Fuel Cells group, Catalonia Institute for Energy Research (IREC), Jardins de les Dones de Negre 1, Barcelona 08930, Spain
[d]Synchrotron Soleil, Gif-Sur-Yvette 91192, France

# Supplementary Information

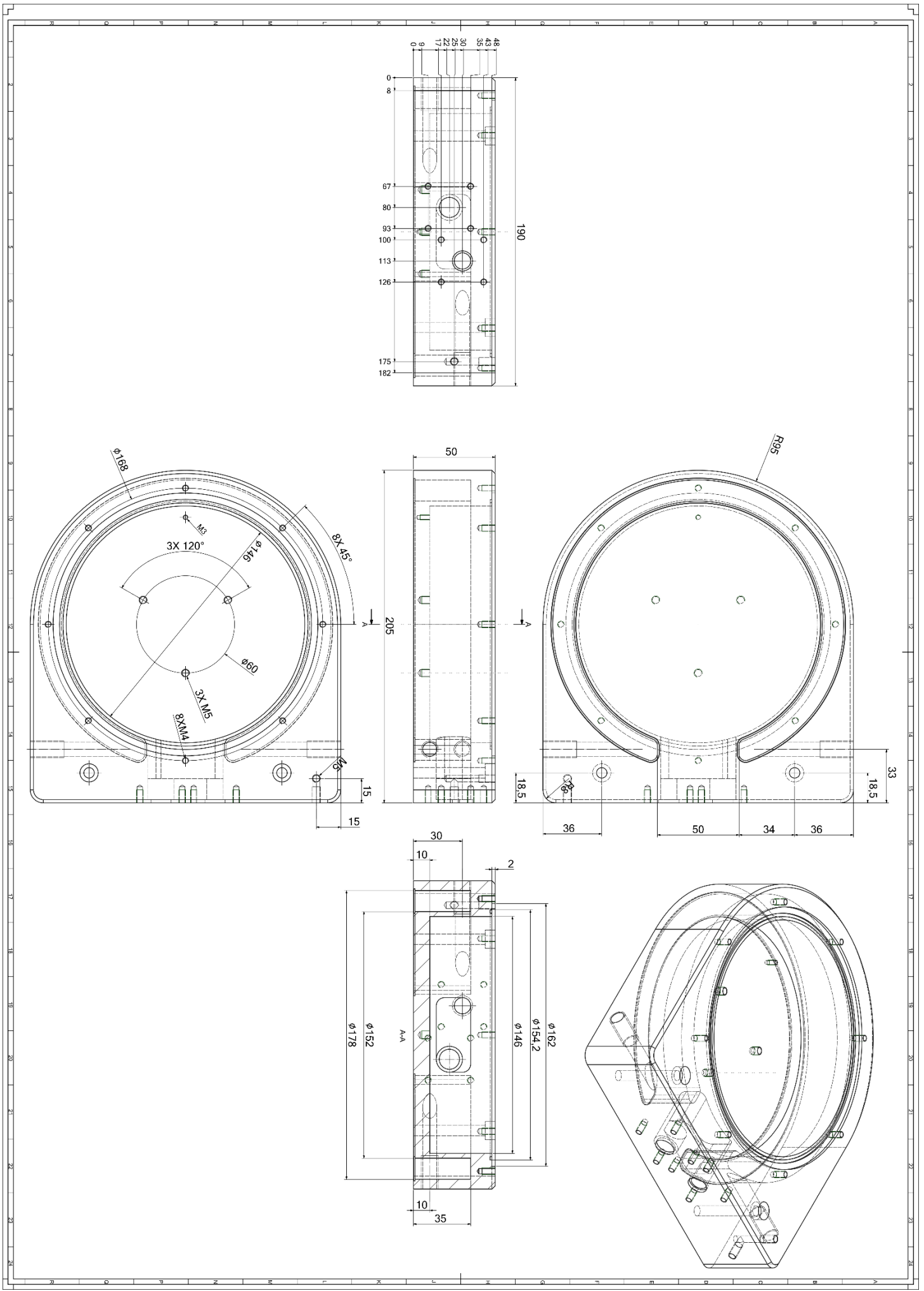

**Figure S1** – Technical drawing with orthographic projections of the furnace support made of stainless steel Type 316L.

**Table S1** – Results of the peak fitting and minimization Matlab routine as platinum lattice parameters with standard deviations for heating and cooling measurements.

| Temperature [°C] | Ramp | Pt Lattice Parameter [Å] |
|---|---|---|
| 20 | Heating | 3.9222 ± 0.0004 |
| 50 | Heating | 3.9232 ± 0.0001 |
| 100 | Heating | 3.9250 ± 0.0003 |
| 150 | Heating | 3.9266 ± 0.0003 |
| 200 | Heating | 3.9287 ± 0.0003 |
| 250 | Heating | 3.9306 ± 0.0002 |
| 300 | Heating | 3.9323 ± 0.0002 |
| 350 | Heating | 3.9343 ± 0.0001 |
| 400 | Heating | 3.9360 ± 0.0001 |
| 450 | Heating | 3.9382 ± 0.0002 |
| 500 | Heating | 3.9402 ± 0.0001 |
| 550 | Heating | 3.9422 ± 0.0001 |
| 600 | Heating | 3.9443 ± 0.0001 |
| 650 | Heating | 3.9465 ± 0.0001 |
| 700 | Heating | 3.9486 ± 0.0001 |
| 750 | Heating | 3.9508 ± 0.0001 |
| 800 | Heating | 3.9532 ± 0.0001 |
| 850 | Heating | 3.9554 ± 0.0001 |
| 600 | Cooling | 3.9446 ± 0.0001 |
| 400 | Cooling | 3.9364 ± 0.0001 |
| 200 | Cooling | 3.9284 ± 0.0001 |
| 50 | Cooling | 3.9227 ± 0.0001 |

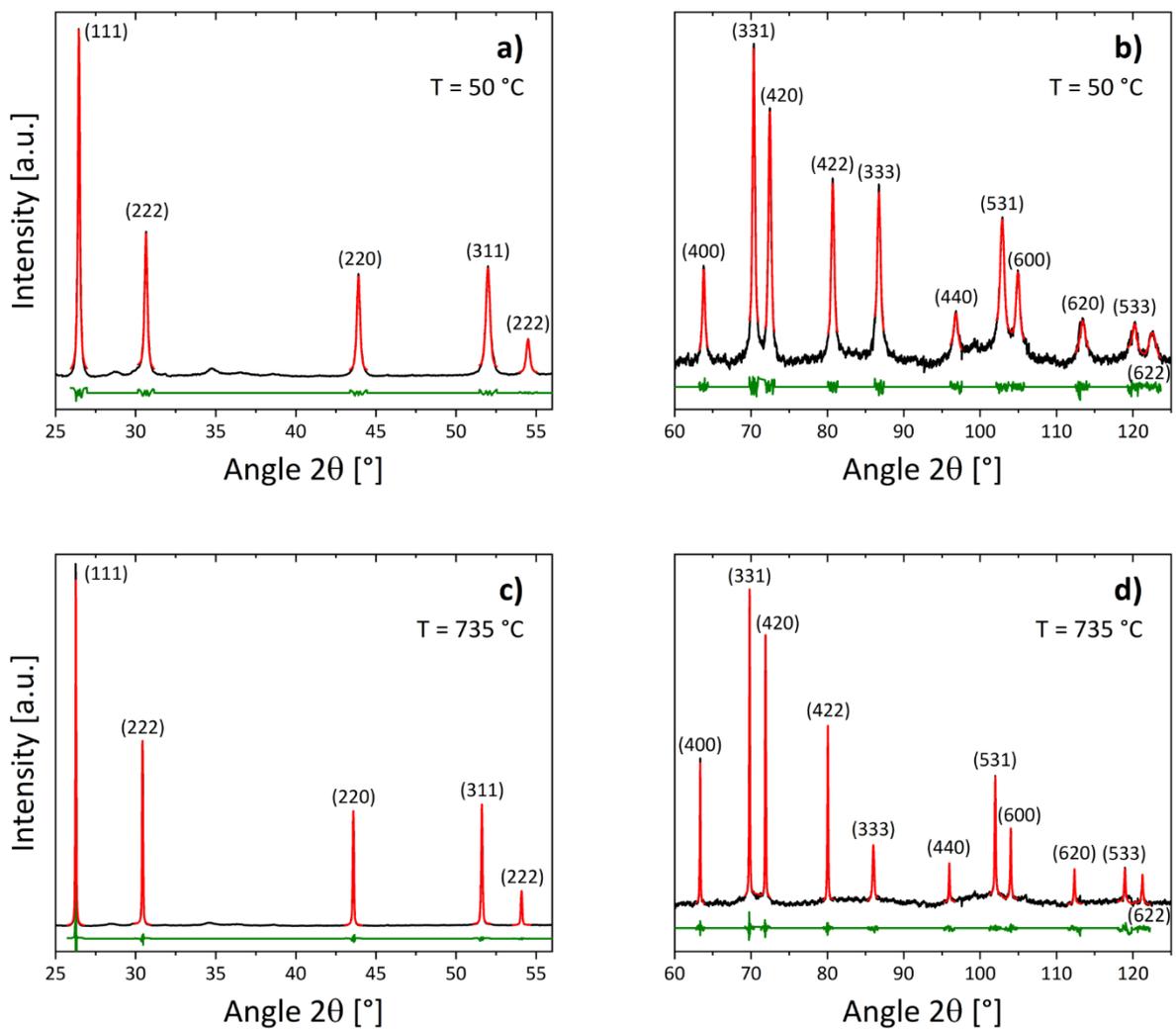

**Figure S2.** XRD diffractograms collected in the center (coordinates: X = 0; Y = 0 mm) of the Pt-coated Si(100) wafer inside the furnace mounted at the DiffAbs beamline of the Soleil synchrotron. The measurements are performed using the PEEK dome at 50 °C (a and b) and 735 °C (c and d) with experimental data (black line), fitted curves (red line), and the residuals (green line). The Miller indices of the first 16 reflections are reported. The scale is different among all 4 panels.

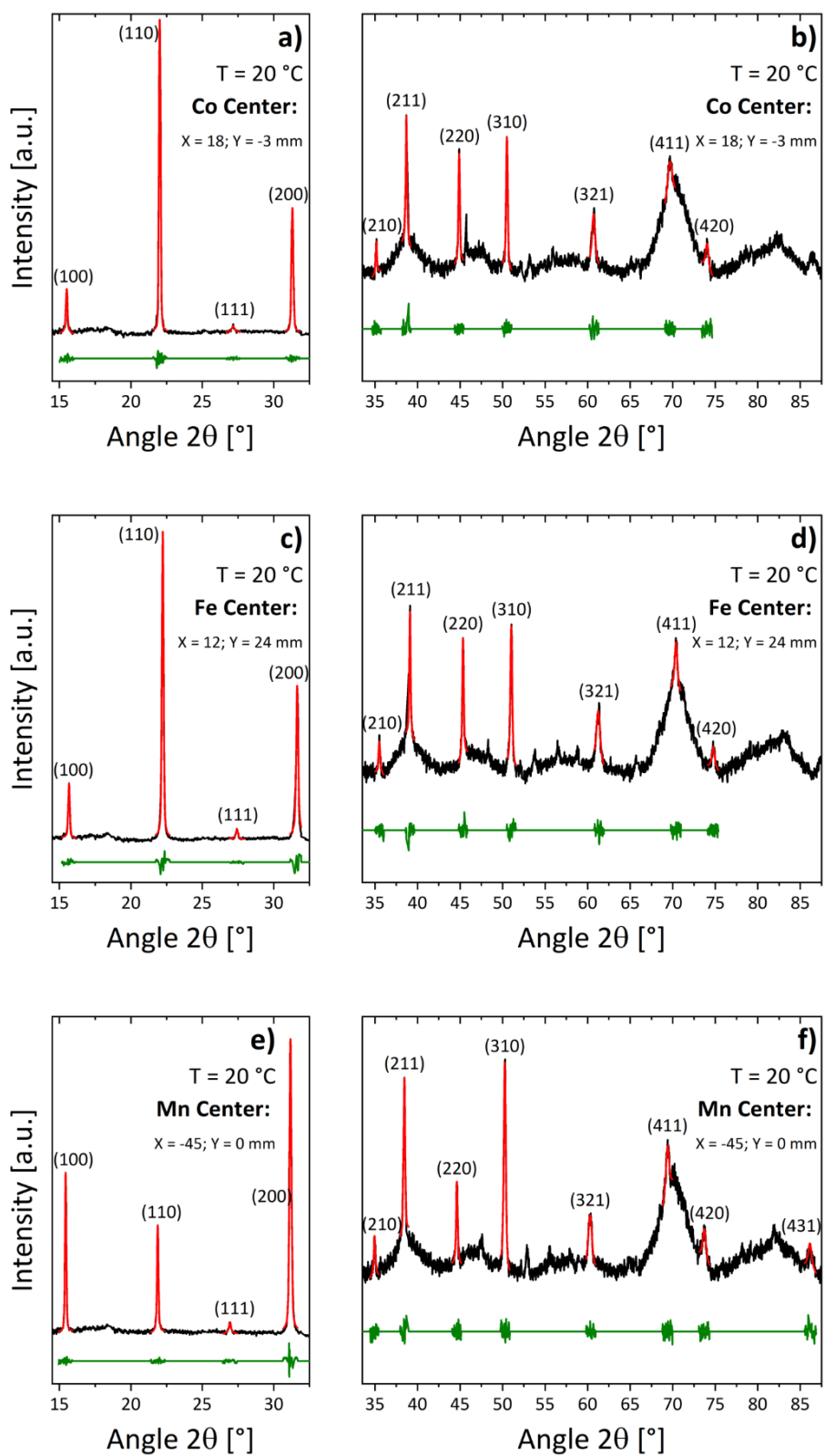

**Figure S3.** XRD diffractograms collected in the Co (a and b, sample coordinates: X = 18; Y = -3 mm), Fe (c and d, sample coordinates: X = 12; Y = 24 mm), and Mn (e and f, sample coordinates: X = -45; Y = 0 mm) of the combinatorial $La_{0.8}Sr_{0.2}Co_{1-x-y}Fe_xMn_yO_{3-\delta}$ (LSCFM) library at the DiffAbs beamline of the Soleil synchrotron at room temperature (20 °C) without the PEEK dome with experimental data (black line), fitted curves (red line), and the residuals (green line). The Miller indices of the first 12 reflections are reported. The scale is different among all 6 panels.

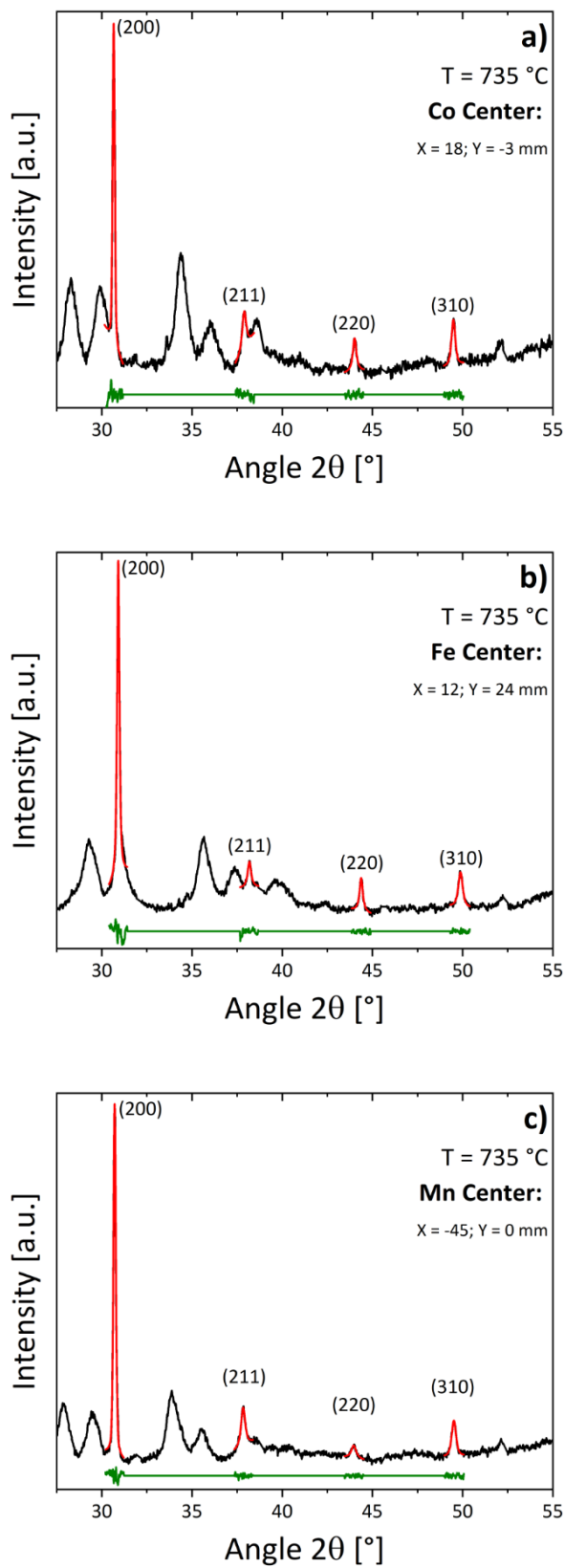

**Figure S4.** XRD diffractograms collected in the Co (a, sample coordinates: X = 18; Y = -3 mm), Fe (b, sample coordinates: X = 12; Y = 24 mm), and Mn (c, sample coordinates: X = -45; Y = 0 mm) of the combinatorial $La_{0.8}Sr_{0.2}Co_{1-x-y}Fe_xMn_yO_{3-\delta}$ (LSCFM) library inside the furnace mounted at the DiffAbs beamline of the Soleil synchrotron. The measurements are performed using the PEEK dome at 735 °C

with experimental data (black line), fitted curves (red line), and the residuals (green line). The Miller indices of the fitted 4 reflections are reported. The scale is different among all 3 panels.

**Table S2** – Results of the in itu XRD high-temperature measurements on the Pt-coated wafer: Pt lattice parameter is expressed in Å, while the temperatures are in °C.

| Setpoint temperature | Thermocouple temperature | Dome temperature | Pt lattice parameter | Minimum temperature | Mean temperature | Maximum temperature |
|---|---|---|---|---|---|---|
| 20  | 20.0  | 20.0  | 3.9219 | 6.6   | 19.0  | 31.0  |
| 50  | 43.6  | 24.7  | 3.9223 | 17.1  | 31.0  | 44.9  |
| 100 | 84.7  | 28.1  | 3.9237 | 54.6  | 70.7  | 83.8  |
| 200 | 168.1 | 37.3  | 3.9268 | 139.3 | 154.3 | 165.1 |
| 250 | 208.9 | 42.5  | 3.9287 | 192.0 | 204.8 | 217.8 |
| 300 | 249.1 | 48.5  | 3.9304 | 239.2 | 251.0 | 265.1 |
| 350 | 288.6 | 52.5  | 3.9321 | 284.0 | 296.0 | 309.0 |
| 400 | 327.2 | 59.3  | 3.9338 | 324.1 | 338.8 | 353.6 |
| 450 | 365.2 | 67.2  | 3.9355 | 357.2 | 381.8 | 398.5 |
| 500 | 402.8 | 71.0  | 3.9371 | 395.7 | 422.2 | 443.1 |
| 550 | 439.3 | 86.3  | 3.9390 | 430.3 | 468.6 | 493.0 |
| 600 | 476.6 | 113.5 | 3.9405 | 467.2 | 505.9 | 530.7 |
| 650 | 509.7 | 107.5 | 3.9421 | 500.7 | 545.3 | 578.1 |
| 700 | 545.7 | 133.3 | 3.9439 | 543.6 | 586.7 | 625.3 |
| 735 | 570.5 | 139.4 | 3.9451 | 571.4 | 616.3 | 652.8 |